\documentclass[12pt,doublespacing]{revtex4-1}

\usepackage[scale=0.8]{geometry}
\usepackage[utf8]{inputenc}
\usepackage[english]{babel}
\usepackage{amssymb}
\usepackage{amsmath}
\usepackage{graphicx}
\usepackage[caption=false]{subfig}
\usepackage{float}
\usepackage[normalem]{ulem}
\usepackage{color}
\usepackage{placeins}
\usepackage{dashbox}
\usepackage{hyperref}
\usepackage[bottom]{footmisc}
\usepackage{cleveref}
\usepackage{movie15}
\usepackage{blindtext}

\definecolor{orange}{RGB}{255,127,0}
\definecolor{purple}{RGB}{195,76,200}
\definecolor{ecol}{RGB}{170,0,136}
\newcommand{\av}[1]{\langle #1 \rangle}

\newcommand\s{$\mathbb{S}^1$~}

\newcommand\cons{\av{\mathrm{Cons}}}
\newcommand{\FigPath}{./Figs}

\hypersetup{
    colorlinks=true,
    linkcolor=black,
    filecolor=black,  
    citecolor=black,    
    urlcolor=black,
    linktoc=all,
} 
\begin{document}

\author{Elisenda Ortiz}
\affiliation{Departament de F\'{i}sica de la Mat\`{e}ria Condensada, Universitat de Barcelona, Mart\'{i} i Franqu\`{e}s 1, 08028 Barcelona, Spain}
\affiliation{Universitat de Barcelona Institute of Complex Systems (UBICS), Universitat de Barcelona, Barcelona, Spain}

\author{M. \'{A}ngeles Serrano}
\email{marian.serrano@ub.edu}
\affiliation{Departament de F\'{i}sica de la Mat\`{e}ria Condensada, Universitat de Barcelona, Mart\'{i} i Franqu\`{e}s 1, 08028 Barcelona, Spain}
\affiliation{Universitat de Barcelona Institute of Complex Systems (UBICS), Universitat de Barcelona, Barcelona, Spain}
\affiliation{ICREA, Pg. Llu\'{i}s Companys 23, E-08010 Barcelona, Spain}

\date{\today}

\title{Multiscale Voter Model on Real Networks}

\begin{abstract}


We introduce the Multiscale Voter Model (MVM) to investigate clan influence at multiple scales---family, neighborhood, political party...---in opinion formation on real complex networks. Clans, consisting of similar nodes, are constructed using a coarse-graining procedure on network embeddings that allows us to control for the length scale of interactions. We ran numerical simulations to monitor the evolution of MVM dynamics in real and synthetic networks, and identified a transition between a final stage of full consensus and one with mixed binary opinions. The transition depends on the scale of the clans and on the strength of their influence. We found that enhancing group diversity promotes consensus while strong kinship yields to metastable clusters of same opinion. The segregated domains, which signal opinion polarization, are discernible as spatial patterns in the hyperbolic embeddings of the networks. Our multiscale framework can be easily applied to other dynamical processes affected by scale and group influence.
\end{abstract}

\maketitle

\section{Introduction}
Opinion dynamics can be modeled using interacting agents in social networks in order to investigate the spreading of attitudes, beliefs, and sentiments in society. In this context, the Voter Model (VM) is an archetypal stochastic nonequilibrium model that gives a standard framework for studying imitation as an underlying mechanism of opinion formation~\cite{Suchecki2005a,castellano2009statistical}. In networks, the small-world property substantially reduces the time to reach consensus in finite systems~\cite{Castellano2003,Vilone2004}, and heterogeneous distributions of the number of neighbors also promote quick agreement~\cite{Sood2008}. Conversely, in real life scenarios we rarely find a large group of individuals easily coming to a consensus on sensitive topics. This dichotomy has motivated generalizations of the VM that include more realistic features such as  zealots, bounded confidence, noise or memory effects~\cite{Redner2019}.

Here, we address this contradiction by introducing the Multiscale Voter Model (MVM), which assumes the decisions of an individual are affected by the viewpoint of its own group. Although group-level information is known to affect behavioural responses in human~\cite{Baumann2020} and even in animal~\cite{Hobson2021} social networks, few models account for it. Among them, there is the q-voter model~\cite{PhysRevE.80.041129}, where an agent takes the opinion of \textit{q} connected neighbors that agree, the majority-vote model~\cite{vilela2020three}, where a node copies the state of the majority of its neighbors, and other models that introduce similar types of non-linearities~\cite{peralta2018analytical}. Other alternatives use multiplex network representations~\cite{Diakonova2016,Amato2017}, or couple individual information exchange with external information fields~\cite{Tsarev2019}. Instead, the MVM relies on the geometric embedding of a network~\cite{boguna2020network} to define clans of similar nodes (not necessarily neighbors) at some specific granularity---family, neighborhood, political party, country---that influence the decision of adopting the opinion of a neighbor. The model interpolates in a natural way between states that reach consensus fast, as in the VM in small-world networks, and frozen disordered states typical of lattices, going through competition between opinion domains.
 
In this contribution, the interpretation of a clan goes beyond the normative meaning of extended family to define a group that has a shared identity based on previous experiences. This definition directly links to topological properties: when a clan is made of individuals who have been in contact for a long time and gone through similar experiences, sociology suggests we should expect a high degree of clustering within the members. This topological insight motivates using similarity distances in geometric embeddings of networks to assign nodes into clans.

\begin{figure}[t]
\centering
\scalebox{1.0}{
  \includegraphics[width=0.5\linewidth]{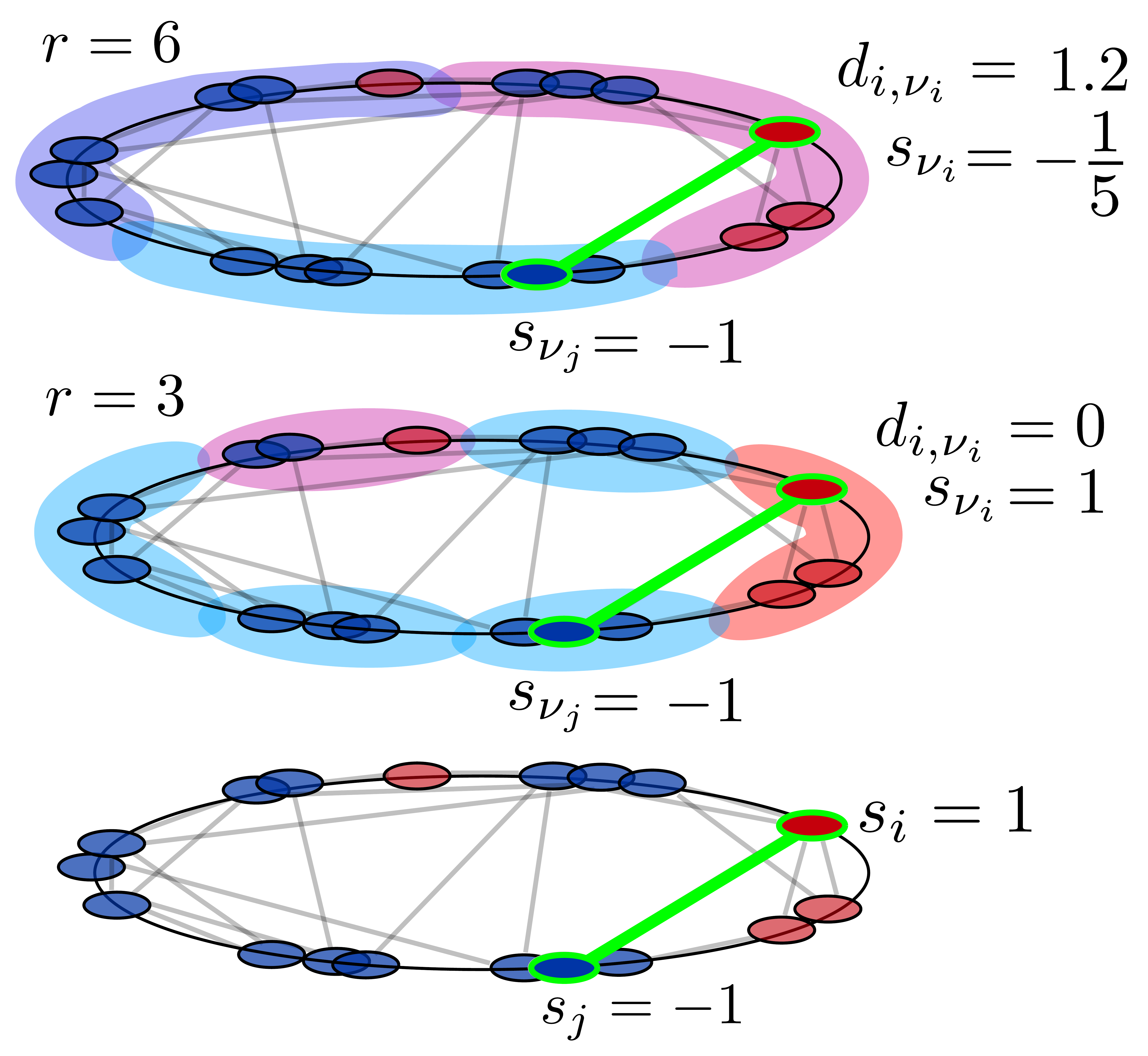}
}
\caption{\textbf{Illustration of the MVM model.} In the network at the bottom, the link between a copycat node $i$ with opinion $s_i=1$ and its neighbor $j$ with opposite opinion $s_j=-1$ is highlighted in green. Similarity clans in the middle ($r=3$) and upper ($r=6$) layers are colored according to their opinion, ranging from -1 (all nodes blue) to 1 (all red) going through mixed composition (purple). In the middle layer, node $i$ is completely aligned with the opinion of its clan $c$, so that when $r=3$ the probability to copy $j$ is low. For clans of size $r=6$ at the top, the distance in opinion between node $i$ and its clan increases and so does the probability to copy $j$.}
      \label{Fig1}
\end{figure}

\section{The Multiscale Voter Model}
In the MVM each node $i$ holds one of two possible opinions $s_i=\{-1,1\}$, and nodes interact by copying the opinion state of a randomly chosen neighbor. To avoid biases induced by heterogeneous degrees, we implemented a link update dynamics where links of the network are selected uniformly at random~\cite{PhysRevE.77.041121}. Then, one of the two linked nodes copies the opinion of its neighbor with a certain probability tuned by the influence of its own clan $c$. For that purpose, we introduce a distance $d_{i,c}=|s_i -s_{c}| \in[0,2]$ between the opinion $s_i$ of node $i$ and that of its clan $s_{c}$, which weights the probability of node $i$ adopting the state of a neighbor,
\begin{equation}
P_{i}= \frac{1}{1+e^{\frac{1}{\lambda}(1-d_{i,c})}}\ .
\label{Eq.MVM-Probability}
\end{equation}

The opinion of clan $c$, to which node $i$ belongs, is continuous in $[-1,1]$ and given by the average $s_{c} = \sum_{l\neq i}s_l/(r-1), l \in c$, where node $i$ is excluded and $r$ indicates the total number of nodes in the group. Parameter $\lambda \in [0,\infty)$ in Eq.\ref{Eq.MVM-Probability} controls the strength of the clan influence, which decreases as $\lambda$ increases. The probability $P_i$ is designed to reflect the tendency of individuals to refrain from adopting behaviours that contradict their group norm. Its Fermi-like functional form is also a popular updating protocol in evolutionary game theory, which constitutes a reference framework for addressing stochasticity in human decision-making processes. $P_i$ is symmetric around $d_{i,c}=1$, so that a node that is very aligned with its environment ($d_{i,c}\rightarrow 0$) has less chances of copying a random neighbor, while the probability increases when the node is not aligned with the opinion of its own clan ($d_{i,c}\rightarrow 2$).
When $\lambda\rightarrow 0$, Eq.~(\ref{Eq.MVM-Probability}) tends to a step-function, which leads the system to frozen disordered states. In the limit $\lambda\rightarrow \infty$, the MVM becomes the VM with a rescaled activation rate that slows the dynamics but eventually leads to consensus. The case with $0<\lambda<\infty$ is akin to introducing heterogeneous and dynamic activation rates dependant upon the states of nodes and their clans, and leads to competition between metastable opinion domains.

The implementation of the MVM relies on clans made of similar nodes detected in network embeddings. To generate the embeddings, we used the tool Mercator~\cite{Garcia2019} which produces a representation of a network in the hyperbolic plane. These embeddings are based on the geometric \s model~\cite{Serrano:2008ga}, where every node $i$ has a popularity-similarity pair of coordinates $(\kappa_i, \theta_i)$. Coordinate $\kappa_i$ is proportional to the node's degree and can be mapped into a radial coordinate $r_i$ in the hyperbolic disc -- where nodes with larger degrees lie closer to the center. The angular coordinate $\theta_i$ designates the node position in a circle where two nodes are more similar to each other when they have a shorter angular separation $\Delta \theta_{ij}$, see Supplemental Material (SM) Fig.S1.
 
Given the embedding of a network, clans are constructed by dividing the similarity circle in angular sectors containing $r$ consecutive nodes (see Fig.~\ref{Fig1}). Notice, when $N$ is odd the last clan will have less than $r$ nodes. This coarse-graining procedure is at the core of the geometric renormalization group~\cite{Garcia2018Renorm}, which unfolds a network into a self-similar shell of layers with decreasing resolution for increasing $r$ and progressively more dominated by long range connections. This means the size $r$ of the similarity clans (SC) allows us to control for observation scale.

Besides, we made random groups (RG), also of equal size $r$, in order to provide a null model to gauge unexpected behaviour. Lastly, groups were constructed based on geometrically detected communities identified by applying the Critical Gap Method (CGM) to network embeddings~\cite{Serrano2012a, Garcia2016Atlas}. This was possible since hyperbolic embeddings of real networks present heterogeneously populated angular regions, which indicate meaningful communities of different sizes, see SM for further details.

\section{MVM simulations}
We simulated the MVM dynamics in real and synthetic networks, starting from a random uniform distribution of states with an initial density $\rho(t_0)=0.5$ of nodes in state $s=1$. Notice that fixing $\rho(t_0)$ implies correlated initial opinions. For a randomly picked link, the algorithm selects with equal probability a node $i$, who then adopts the opinion of node $j$ at the other end with $P_i$ given by Eq.~(\ref{Eq.MVM-Probability}). At each simulation step, time is advanced by $\Delta t = (1/E)$, where $E$ is the number of links in the network. Although finite size effects will eventually lead the dynamics to an absorbing consensus state, some realizations can be extremely long-lived. Hence, we set a cutoff time $t_{\mathrm{c}}$, see Tab.~S1 in SM. We measured the level of consensus in the network  $\cons=\av{|\rho(t_{\mathrm{c}})-0.5|/0.5}$, where the average is over independent realizations, and computed the fluctuations $\chi=(\av{\mathrm{Cons^2}}-\cons^2)/\cons$, where we chose the normalization factor following Ref.~\cite{Colomer2014}. Finally, we also evaluated the survival probability $S$, measuring the fraction of realizations that did not reach consensus at $t_{\mathrm{c}}$, to elucidate how individual realizations contribute to the average level of consensus.

\subsection{Results for real networks}
We considered four data sets from different domains where clans find a natural interpretation: a New Zealand Members of Parliament political network~\cite{Curran2018} (NZ-MPs), a Facebook friendship network~\cite{Traud2011} (Facebook), a social proximity network of bottlenose dolphins~\cite{Gazda2015} (Dolphins), and the World Trade Web (WTW)~\cite{Garcia2016Atlas}.

\begin{figure}[t]
\centering
\scalebox{1.0}{
  \includegraphics[width=0.5\linewidth]{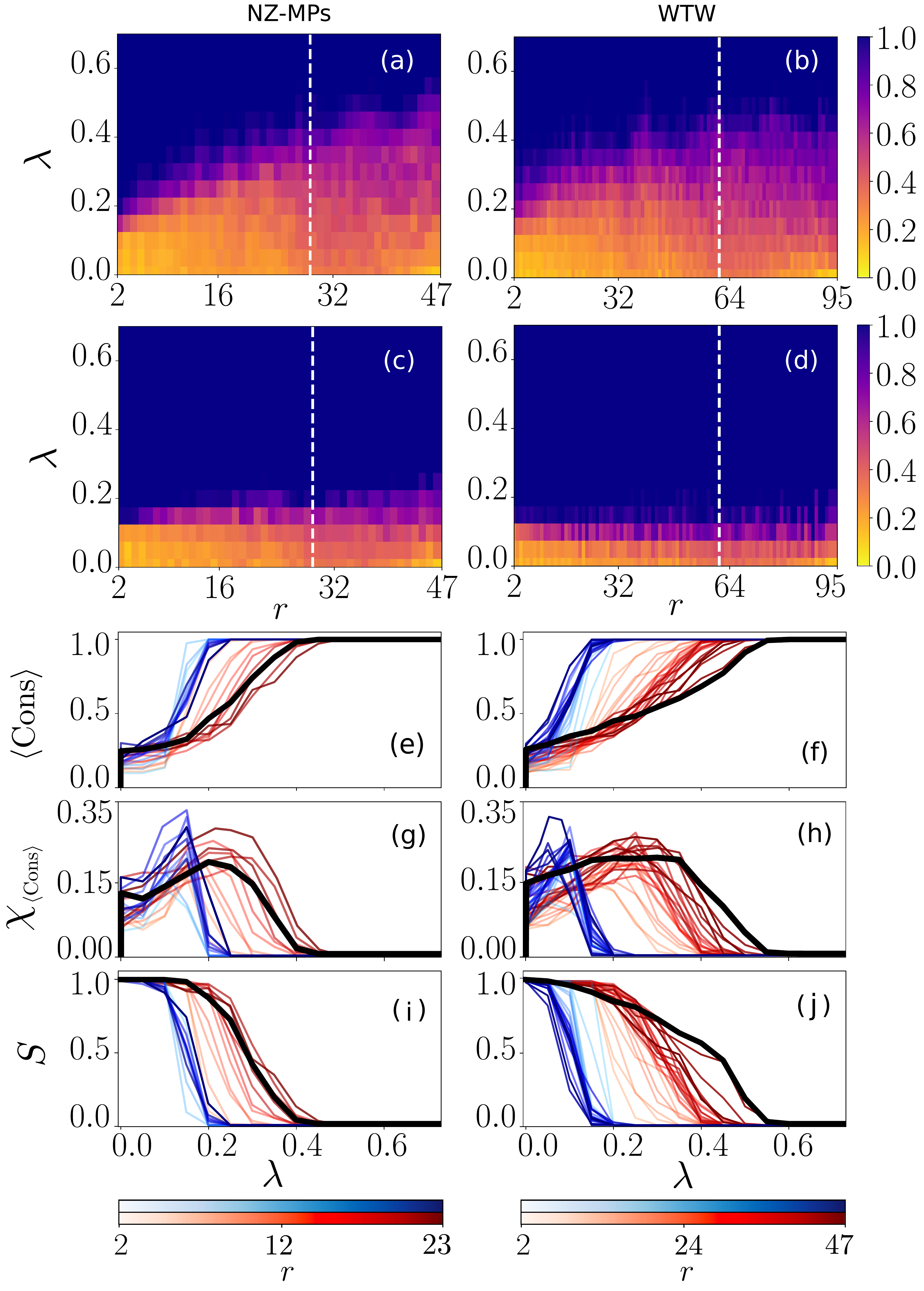}
}
\caption{\textbf{MVM consensus in real networks.} Consensus heatmaps using SC \textbf{(a)-(b)} and RG \textbf{(c)-(d)} over 100 realizations. Parameter space confined in $r\in [2,N/4]$ and $\lambda \in [10^{-5}, 0.7]$. A white dashed line across the heatmap denotes the size $r_c$ of the largest community detected via the CGM. In plots (e)-(j) red curves correspond to SC and blue ones to RG. Darker tones indicate larger group sizes in both cases. Solid black lines denote groups corresponding to communities. Average consensus $\cons$ \textbf{(e)-(f)}, fluctuations of average consensus level \textbf{(g)-(h)}, and survival probability \textbf{(i)-(j)} against strength of group-influence $\lambda$ for several group sizes $r$.}
      \label{Fig3}
\end{figure}

Results for the four real networks are qualitatively similar, NZ-MPs and WTW are shown in Fig.~\ref{Fig3}, and Facebook and Dolphins in Figs.~S2-S3. In Fig.~\ref{Fig3}(a)-(b), we show consensus heatmaps in the $(\lambda, r)$ configuration space for the MVM dynamics using similarity groups SC, and the control case of random groups RG in Fig.~\ref{Fig3}(c)-(d). For comparability, $r_{\mathrm{max}}$ is chosen as the group size that divides a network in two portions of $N/2$. The heatmaps show areas of low, moderate and high levels of final agreement, with intermediate consensus values more predominant within $\lambda\in [0.15-0.5]$ as $r$ grows. In contrast, for smaller clans $r\lesssim 20$, intermediate levels of consensus are more difficult to sustain in all real networks, specially in the region $\lambda \in [0.25-0.30]$. This indicates that clan influence at smaller scales dictates more drastically whether the system evolves towards global agreement or not. Furthermore, when groups are random, all real networks display a transition centered at $\lambda_{\mathrm{crit}}\approx 0.15$ independent of scale $r$, see also Fig.~S1. This means that mixed opinion configurations are invariably less stable over time when groups do not capture actual similarities.

In Fig.~\ref{Fig3}(e)-(f), we show a projection of the level of consensus against $\lambda$. All networks show a more abrupt transition for RG than for SC, and SC slopes show significant variation with $r$. The fluctuations $\chi$ in Fig.~\ref{Fig3}(g)-(h) show maxima around $\lambda_{\mathrm{crit}}(r)<0.2$ for RG, while for SC peaks appear at higher values $\lambda_{\mathrm{crit}}(r)>0.2$ and present lower maxima in all cases. The survival probability $S$ at the bottom of Fig.~\ref{Fig3}, shows a decreasing trend with $\lambda$ which demonstrates agreement is more easily achieved as group influence is dissolved. However, the decay is more abrupt for RG curves, which indicates that at $\lambda_{\mathrm{crit}}(r)$ the networks are in very low agreement configurations, but as soon as $\lambda >\lambda_{\mathrm{crit}}(r)$ most simulation realizations reach full consensus before the cutoff time. Simulations with smaller $r$ in Fig.~\ref{Fig3}(i)-(j) decay fast and continuously to 0 while the process is slower and less monotonous for larger values of $r$, see Fig.~S4. Besides, compared to RG, SC curves show less overlap and are shifted towards larger values of $\lambda$, indicating a more progressive transition to consensus demanding more freedom from the group influence.

Finally, we simulated the MVM dynamics using groups corresponding to geometric communities detected via the Critical Gap Method~\cite{Garcia2016Atlas}. Results of running the MVM dynamics for CGM communities are reported using black solid lines on Fig.~\ref{Fig3}(e)-(j) and Figs.~S3-S4. Interestingly, when examining $\cons$, $\chi$, and $S$, we identify a pattern that holds across networks despite their different nature, number of communities $n_c$, and size of the largest community $r_c$ (see Table S1). This is, the results for CGM communities follow approximately the trend of the results for SC curves of $r=r_{c}$. This means that the largest community of the network is effectively ruling the evolution and eventual outcome of the MVM dynamics.

\subsection{Results for synthetic \s networks}
Additionally, simulations on \s synthetic networks allowed us to corroborate the above results and estimate the impact of specific topological features in the final stage of the dynamics. We generated  \s synthetic networks of sizes $N=1000, 5000$ nodes and realistic parameters $\gamma$ and $\beta$, that control the scale-freeness of the degree distribution and the mean clustering coefficient, respectively . Consensus heatmaps, fluctuations, and the survival probability are shown in Fig.~S5. As observed for real networks, SC clans undergo a progressive transition from low to high consensus within a range of $\lambda$ values particular of each network and dependent on the group scale $r$, while RG groups exhibit a sharp transition localized at $\lambda_{\mathrm{crit}}\lesssim 0.15$ with unanimity dominating the majority of the phase space. We confirm that a minor increase in $\lambda$ for small SC clans can lessen the strength of group influence enough to suddenly push the system towards fast consensus as happens with RG. Conversely, for larger SC clans, a higher rate of opinion exchange allows to sustain more intermediate levels of global consensus for a wider range of $\lambda$, see Fig.~S5(g)-(h). In summary, we validated the results obtained for real networks and found that unanimity is usually harder to achieve as network topology becomes more homogeneous and clustered (increasing $\gamma$ and $\beta$), see Figs.~S6-S7.

\begin{figure}[t]
\centering
\scalebox{1.0}{
  \includegraphics[width=0.5\linewidth]{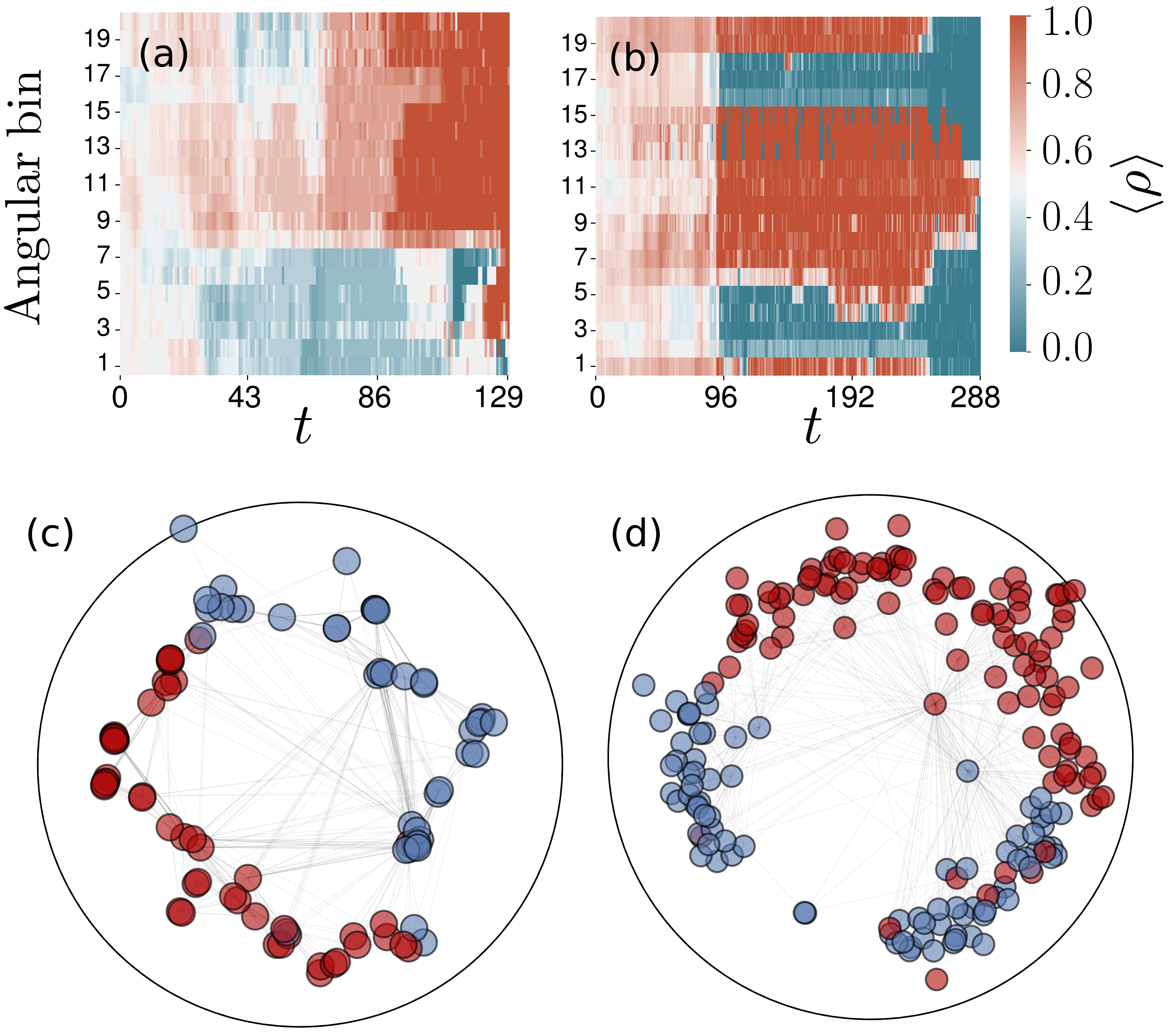}
}
\caption{\textbf{Geometric opinion domains} of the NZ-MPs (left colum) and the WTW (right colum) networks. \textbf{(a)-(b)} Time evolution of average density of nodes in state $s=1$ $\av \rho$ in 20 equally sized bins of the angular coordinate $\theta$ ($r=10$, $\lambda=0.45$). \textbf{(c)-(d)} Hyperbolic maps of a snapshot of the MVM dynamics using SC ($r=10$, $\lambda=0.45$). Two angular domains of different opinion are visible. Red for state $s=1$ and blue for state $s=-1$. 
}
\label{Fig4}
\end{figure}

\section{Opinion domains}
Following, we show that similarity clans in the MVM dynamics trigger the formation of meta-stable clusters of homogeneous opinion in similarity space, which prevent rapid collapse into consensus. Fig.~\ref{Fig4}(a)-(b) show the evolution of the average density $\av \rho$ of nodes in state $s=1$ in equally sized angular bins for the NZ-MPs and WTW, respectively. Initially, two opinions, $s=\{-1,1\}$, were homogeneously spread across the similarity space of both networks. As time passed, $\av\rho$ increased in the NZ-MPs on a wide region (red area), while the opposite opinion (blue) prevailed in the remaining  angular space. For the WTW, two clusters of $s=1$ (notice continuous boundary conditions) remained for most of the simulation separated by two other clusters of opposite opinion $s=-1$, which eventually joined ends making $\av\rho=0$.
See also the evolution of the average group opinion over time for similarity groups in Fig.~S8. For both networks, we observe a remarkable correspondence between angular proximity of clans and the establishment of polarized opinions.

In SM Fig.~S9, we report WTW results for various clan sizes $r$ while fixing $\lambda=0.6$. We find that opposite opinions tend to develop in separate angular regions independently of $r$ until one colonizes the space of the other or a big fluctuation drives the system to sudden full consensus. In general, we rarely observe more than one angular domain per opinion such as displayed in Fig.~\ref{Fig4}(b). Besides, we fixed the group size to $r=10$ for both datasets, expanded the range of $\lambda$ values used to simulate the evolution of $\av \rho$ and compared SC and RG prescriptions in Figs.~S10-S11. Importantly, we found that random groups do not sustain geometric domains over time. Instead, a higher or lower $\av \rho$ may alternate at times but always spreading homogeneously across all angular bins. Besides, all RG simulations last significantly shorter on average than the ones with similarity clans. We, thus, confirm that similarity between nodes is key to support metastable opinion clusters.

In Figs.~\ref{Fig4}(c)-(d), the hyperbolic maps showcase the spatial distribution of opinions in network snapshots for a single realization of the dynamics. Akin to domain formation in lattice topologies, we visualize the emergence of clusters of homogeneous opinion along the angular dimension of the hyperbolic disc. Furthermore, we provide two animations comparing the MVM temporal evolution of node states in the hyperbolic maps of the WTW, under SC and RG respectively~\cite{gitMVM}. The animation for SC clearly features two spatial clusters sustained over time. We note also that nodes alternating state most frequently are positioned at the borders of the two adjoining opinion domains. On the contrary, the animation using RG does not exhibit any opinion segregation of node states along the circle and the dynamics evolves without nodes being more active in any particular region.

\section{Analytic results}
Finally, we were able to obtain equations for the time evolution of a clan using mean field approximations. At the microscopic level, the dynamical state of a node $i$ after a time interval $dt$ is
\begin{equation}
s_i(t+dt)=s_i(t)(1-\mu_i(dt))+\mu_i(dt) \xi_i,
\label{eq1}
\end{equation} 
where $\mu_i(dt)$ and $\xi_i$ are two independent binary stochastic variables that define the transitions between states. The variable $\mu_i(dt)$ takes the value 1 or 0 depending on whether $i$ was activated for a change or not. In case node $i$ was activated, $\xi_i$ takes the value 1 if $i$ copies a neighbor with state 1, and -1 otherwise. Thus, $\mu_i(dt)$ and $\xi_i$ have probability distributions 
\begin{eqnarray}
P(\mu_i(dt))= P_{i}(t) k_i dt \delta_{\mu_i,1}+(1-P_{i}(t) k_i dt )\delta_{\mu_i,0}\\
P(\xi_i)=\Phi_i(t)/k_i \delta_{\xi_i,1}+(1-\Phi_i(t)/k_i) \delta_{\xi_i,-1}
\label{eq2}
\end{eqnarray}
with $\Phi_i(t)=\sum_j a_{ij}(s_j(t)+1)/2$. A factor $\omega/(N\left<k \right>)$ has been reabsorbed in the definition of $dt$, where $\omega$ is the constant rate for the occurrence of events, which follow an independent Poisson process for each node. These equations together with Eq.~(\ref{eq1}) give a complete but untractable description of the evolution of the system. To gain some understanding we perform an ensemble average over realizations conditioned to the state of the system at time $t$ and obtain the mean field equation for the evolution of the state of a node
\begin{equation}
\dot{\overline{s}}_i(t)=\overline{P}_i(t)[-\overline{s}_i(t)k_i+2\overline{\Phi}_i(t)-k_i],
\label{simean}
\end{equation} 
where we have made the assumption that the averages of the different quantities can be taken as independent. The expressions above are valid for any network topology.

Next, we consider a clan and use Eq.~(\ref{simean}) to obtain the evolution of its state $\overline{s}_c$ by averaging over members of the clan, 
\begin{equation}
\dot{\overline{s}}_{c}=  \frac{1}{1+e^{\frac{1}{\lambda}} }[k_c (\overline{s}_c +1) + \frac{2}{r}\sum_{l \in c} \overline{\Phi}_l ],
\end{equation} 
where $k_c$ is the average degree of nodes in the clan and we have dropped the time dependence to ease the notation. 

To analyze the stability of a clan, we assume the most extreme situation in which all the nodes in the clan hold the same opinion (state 1), which is the opposite to that of the rest of the network (state -1), and we find
\begin{equation}
\dot{\overline{s}}_{c}=-\frac{1}{1+e^{\frac{1}{\lambda}}} k_c^{ext}[  \overline{s}_c +1] \equiv -F(\lambda,k_c^{ext})[  \overline{s}_c +1],
\label{eqsc}
\end{equation} 
where $k_c^{ext}$ is the average degree of the nodes in the clan restricted to connected neighbors outside the clan. Eq.~(\ref{eqsc}) can be used to estimate the time until the clan aligns with the outside opinion. The solution $\overline{s}_{c}=2 e^{-Ft}-1$ has a characteristic time $1/F$, which depends on the strength of the clan influence $\lambda$ and on the strength of external connections of nodes in the clan, such that the exponential decay to the external state is faster when a clan is more connected with the environment or has a weaker clan influence on its members.

\section{Conclusions}
We have shown that MVM dynamics on real and synthetic networks can reach mixed binary opinions or full consensus depending on the scale of the groups and on the strength of their influence. Specifically, bigger similarity clans can sustain for longer mixed opinion configurations while a large number of small clans yields a more abrupt transition between very low and very high levels of consensus as group influence diminishes. This behaviour is also found for random groups of any size. Beyond group scale and strength of influence, group composition radically affects the ease of reaching consensus. Although large differences in backgrounds and perspectives might be expected to contribute to gridlock, we found that the dynamics typically survived longer without reaching global agreement when groups consisted of affine nodes. This is due to the formation of metastable domains of same opinion, which create visible spatial patterns in the angular dimension of the hyperbolic maps of networks. On the contrary, when groups where randomized the opposite was true. This indicates that group diversity can help promote global agreement by reducing friction between sectors of like-minded individuals that pull in opposite directions. Indeed, real observations support the ability of diverse interdisciplinary teams to operate smoothly~\cite{Balagna2020}.

Consensus is, thus, easier within more diverse groups, and diversity can be achieved either by partitioning or mixing the groups, which helps explain why we do not observe that big structured populations easily come to a full consensus in the real world. However, lack of consensus is not always detrimental in society and the coexistence of a plurality of opinions can also be beneficial~\cite{McKinseyreport}. Consequently, an interesting research venue would be to investigate whether mechanistic behavioural rules that help preserve some degree of diversity pose significant advantages to the system's efficient organization, thus suggesting a possible evolutionary origin.

Our multiscale framework can be exported to other dynamical processes where scale and group influence may have a role. For instance, to understand how social acceptance is modified depending on the backup tribe of the influencer. Another possibility is to include multiscale zealots to mimic political parties with stringent ideologies, or add multiple discrete opinions. At the same time, a complete characterization of the MVM model, including the nature of the observed transition and the existence of conserved quantities, would be interesting and remains for future work.

We thank Alex Arcas for a preliminar related exploratory study and Mari\'an Bogu\~{n\'a} for helpful comments. We acknowledge support from the Agencia Estatal de Investigaci\'on of Spain project number PID2019-106290GB- C22/AEI/10.13039/501100011033.

\bibliographystyle{naturemag}
\bibliography{ref_full.bib}

\begin{thebibliography}{10}
\expandafter\ifx\csname url\endcsname\relax
  \def\url#1{\texttt{#1}}\fi
\expandafter\ifx\csname urlprefix\endcsname\relax\def\urlprefix{URL }\fi
\providecommand{\bibinfo}[2]{#2}
\providecommand{\eprint}[2][]{\url{#2}}

\bibitem{Suchecki2005a}
\bibinfo{author}{Suchecki, K.}, \bibinfo{author}{Egu{\'i}luz, V.~M.} \&
  \bibinfo{author}{Miguel, M.~S.}
\newblock \bibinfo{title}{Voter model dynamics in complex networks: Role of
  dimensionality, disorder, and degree distribution.}
\newblock \emph{\bibinfo{journal}{Physical Review E}}
  \textbf{\bibinfo{volume}{72}}, \bibinfo{pages}{036132}
  (\bibinfo{year}{2005}).

\bibitem{castellano2009statistical}
\bibinfo{author}{Castellano, C.}, \bibinfo{author}{Fortunato, S.} \&
  \bibinfo{author}{Loreto, V.}
\newblock \bibinfo{title}{Statistical physics of social dynamics}.
\newblock \emph{\bibinfo{journal}{Reviews of modern physics}}
  \textbf{\bibinfo{volume}{81}}, \bibinfo{pages}{591} (\bibinfo{year}{2009}).

\bibitem{Castellano2003}
\bibinfo{author}{Castellano, C.}, \bibinfo{author}{Vilone, D.} \&
  \bibinfo{author}{Vespignani, A.}
\newblock \bibinfo{title}{Incomplete ordering of the voter model on small-world
  networks}.
\newblock \emph{\bibinfo{journal}{EPL}} \textbf{\bibinfo{volume}{63}},
  \bibinfo{pages}{153} (\bibinfo{year}{2003}).

\bibitem{Vilone2004}
\bibinfo{author}{Vilone, D.} \& \bibinfo{author}{Castellano, C.}
\newblock \bibinfo{title}{Solution of voter model dynamics on annealed
  small-world networks}.
\newblock \emph{\bibinfo{journal}{Phys Rev E}} \textbf{\bibinfo{volume}{69}},
  \bibinfo{pages}{016109} (\bibinfo{year}{2004}).

\bibitem{Sood2008}
\bibinfo{author}{Sood, V.}, \bibinfo{author}{Antal, T.} \&
  \bibinfo{author}{Redner, S.}
\newblock \bibinfo{title}{Voter models on heterogeneous networks}.
\newblock \emph{\bibinfo{journal}{Phys Rev E}} \textbf{\bibinfo{volume}{77}},
  \bibinfo{pages}{041121} (\bibinfo{year}{2008}).

\bibitem{Redner2019}
\bibinfo{author}{Redner, S.}
\newblock \bibinfo{title}{Reality-inspired voter models: A mini-review}.
\newblock \emph{\bibinfo{journal}{Comptes Rendus Physique}}
  \textbf{\bibinfo{volume}{20}}, \bibinfo{pages}{275--292}
  (\bibinfo{year}{2019}).

\bibitem{Baumann2020}
\bibinfo{author}{Baumann, F.}, \bibinfo{author}{Lorenz-Spreen, P.},
  \bibinfo{author}{Sokolov, I.} \& \bibinfo{author}{Starnini, M.}
\newblock \bibinfo{title}{Modeling echo chambers and polarization dynamics in
  social networks}.
\newblock \emph{\bibinfo{journal}{Phys Rev Lett}}
  \textbf{\bibinfo{volume}{124}}, \bibinfo{pages}{048301}
  (\bibinfo{year}{2020}).

\bibitem{Hobson2021}
\bibinfo{author}{Hobson, E.}, \bibinfo{author}{M{\o}nster, D.} \&
  \bibinfo{author}{DeDeo, S.}
\newblock \bibinfo{title}{Aggression heuristics underlie animal dominance
  hierarchies and provide evidence of group-level social information}.
\newblock \emph{\bibinfo{journal}{PNAS}} \textbf{\bibinfo{volume}{9}},
  \bibinfo{pages}{118} (\bibinfo{year}{2021}).

\bibitem{PhysRevE.80.041129}
\bibinfo{author}{Castellano, C.}, \bibinfo{author}{Mu\~noz, M.~A.} \&
  \bibinfo{author}{Pastor-Satorras, R.}
\newblock \bibinfo{title}{Nonlinear $q$-voter model}.
\newblock \emph{\bibinfo{journal}{Phys. Rev. E}} \textbf{\bibinfo{volume}{80}},
  \bibinfo{pages}{041129} (\bibinfo{year}{2009}).

\bibitem{vilela2020three}
\bibinfo{author}{Vilela, A.~L.} \emph{et~al.}
\newblock \bibinfo{title}{Three-state majority-vote model on scale-free
  networks and the unitary relation for critical exponents}.
\newblock \emph{\bibinfo{journal}{Scientific Reports}}
  \textbf{\bibinfo{volume}{10}}, \bibinfo{pages}{1--11} (\bibinfo{year}{2020}).

\bibitem{peralta2018analytical}
\bibinfo{author}{Peralta, A.~F.}, \bibinfo{author}{Carro, A.},
  \bibinfo{author}{San~Miguel, M.} \& \bibinfo{author}{Toral, R.}
\newblock \bibinfo{title}{Analytical and numerical study of the non-linear
  noisy voter model on complex networks}.
\newblock \emph{\bibinfo{journal}{Chaos: An Interdisciplinary Journal of
  Nonlinear Science}} \textbf{\bibinfo{volume}{28}}, \bibinfo{pages}{075516}
  (\bibinfo{year}{2018}).

\bibitem{Diakonova2016}
\bibinfo{author}{Diakonova, M.}, \bibinfo{author}{Nicosia, V.},
  \bibinfo{author}{Latora, V.} \& \bibinfo{author}{San-Miguel, M.}
\newblock \bibinfo{title}{Irreducibility of multilayer network dynamics: the
  case of the voter model}.
\newblock \emph{\bibinfo{journal}{New J Phys}} \textbf{\bibinfo{volume}{18}},
  \bibinfo{pages}{023010} (\bibinfo{year}{2016}).

\bibitem{Amato2017}
\bibinfo{author}{Amato, N., R.~Kouvaris}, \bibinfo{author}{San-Miguel, M.} \&
  \bibinfo{author}{D\'{i}az-Guilera, A.}
\newblock \bibinfo{title}{Opinion competition dynamics on multiplex networks}.
\newblock \emph{\bibinfo{journal}{New J Phys}} \textbf{\bibinfo{volume}{19}},
  \bibinfo{pages}{123019} (\bibinfo{year}{2017}).

\bibitem{Tsarev2019}
\bibinfo{author}{Tsarev, D.}, \bibinfo{author}{Trofimova, A.},
  \bibinfo{author}{Alodjants, A.} \& \bibinfo{author}{Khrennikov, A.}
\newblock \bibinfo{title}{Phase transitions, collective emotions and
  decision-making problem in heterogeneous social systems}.
\newblock \emph{\bibinfo{journal}{Sci Rep}} \textbf{\bibinfo{volume}{9}},
  \bibinfo{pages}{18039} (\bibinfo{year}{2019}).

\bibitem{boguna2020network}
\bibinfo{author}{Bogu{\~{n}}{\'a}, M.} \emph{et~al.}
\newblock \bibinfo{title}{Network geometry}.
\newblock \emph{\bibinfo{journal}{Nature Reviews Physics}}
  \textbf{\bibinfo{volume}{3}}, \bibinfo{pages}{114--135}
  (\bibinfo{year}{2021}).

\bibitem{PhysRevE.77.041121}
\bibinfo{author}{Sood, V.}, \bibinfo{author}{Antal, T.} \&
  \bibinfo{author}{Redner, S.}
\newblock \bibinfo{title}{Voter models on heterogeneous networks}.
\newblock \emph{\bibinfo{journal}{Phys. Rev. E}} \textbf{\bibinfo{volume}{77}},
  \bibinfo{pages}{041121} (\bibinfo{year}{2008}).
\newblock \urlprefix\url{https://link.aps.org/doi/10.1103/PhysRevE.77.041121}.

\bibitem{Garcia2019}
\bibinfo{author}{Garc{\'{\i}}a-P{\'{e}}rez, G.}, \bibinfo{author}{Allard, A.},
  \bibinfo{author}{Serrano, M.} \& \bibinfo{author}{Bogu{\~{n}}{\'{a}}, M.}
\newblock \bibinfo{title}{Mercator: uncovering faithful hyperbolic embeddings
  of complex networks}.
\newblock \emph{\bibinfo{journal}{New J Phys}} \textbf{\bibinfo{volume}{21}},
  \bibinfo{pages}{123033} (\bibinfo{year}{2019}).

\bibitem{Serrano:2008ga}
\bibinfo{author}{Serrano, M.~{\'{A}}.}, \bibinfo{author}{Krioukov, D.} \&
  \bibinfo{author}{Bogu{\~{n}}{\'{a}}, M.}
\newblock \bibinfo{title}{Self-similarity of complex networks and hidden metric
  spaces}.
\newblock \emph{\bibinfo{journal}{Phys Rev Lett}}
  \textbf{\bibinfo{volume}{100}}, \bibinfo{pages}{078701}
  (\bibinfo{year}{2008}).

\bibitem{Garcia2018Renorm}
\bibinfo{author}{Garc{\'\i}a-P{\'e}rez, G.}, \bibinfo{author}{Bogu{\~n}{\'a},
  M.} \& \bibinfo{author}{Serrano, M.~{\'A}.}
\newblock \bibinfo{title}{Multiscale unfolding of real networks by geometric
  renormalization}.
\newblock \emph{\bibinfo{journal}{Nature Physics}}
  \textbf{\bibinfo{volume}{14}}, \bibinfo{pages}{583--589}
  (\bibinfo{year}{2018}).

\bibitem{Serrano2012a}
\bibinfo{author}{Serrano, M.~{\'A}.}, \bibinfo{author}{Bogu{\~n}{\'a}, M.} \&
  \bibinfo{author}{Sagues, F.}
\newblock \bibinfo{title}{Uncovering the hidden geometry behind metabolic
  networks}.
\newblock \emph{\bibinfo{journal}{Mol. BioSyst.}} \textbf{\bibinfo{volume}{8}},
  \bibinfo{pages}{843--850} (\bibinfo{year}{2012}).

\bibitem{Garcia2016Atlas}
\bibinfo{author}{Garc{\'\i}a-P{\'e}rez, G.}, \bibinfo{author}{Bogu{\~n}{\'a},
  A.~M., Allard} \& \bibinfo{author}{Serrano, M.~{\'A}.}
\newblock \bibinfo{title}{The hidden hyperbolic geometry of international
  trade: World trade atlas 1870-2013}.
\newblock \emph{\bibinfo{journal}{Sci. Rep.}} \textbf{\bibinfo{volume}{6}},
  \bibinfo{pages}{33441} (\bibinfo{year}{2016}).

\bibitem{Colomer2014}
\bibinfo{author}{Colomer-de Sim\'{o}n, P.} \& \bibinfo{author}{Bogu\~{n}\'{a},
  M.}
\newblock \bibinfo{title}{Double percolation phase transition in clustered
  complex networks}.
\newblock \emph{\bibinfo{journal}{PRX}} \textbf{\bibinfo{volume}{4}},
  \bibinfo{pages}{041020} (\bibinfo{year}{2014}).

\bibitem{Curran2018}
\bibinfo{author}{Curran, B.}, \bibinfo{author}{Higham, K.},
  \bibinfo{author}{Ortiz, E.} \& \bibinfo{author}{Vasques-Filho, D.}
\newblock \bibinfo{title}{Look who’s talking: Two-mode networks as
  representations of a topic model of new zealand parliamentary speeches}.
\newblock \emph{\bibinfo{journal}{PloS one}} \textbf{\bibinfo{volume}{13}},
  \bibinfo{pages}{e0199072} (\bibinfo{year}{2018}).

\bibitem{Traud2011}
\bibinfo{author}{Traud, A.}, \bibinfo{author}{Kelsic, E.},
  \bibinfo{author}{Mucha, P.} \& \bibinfo{author}{Porter, M.}
\newblock \bibinfo{title}{Comparing community structure to characteristics in
  online collegiate social networks}.
\newblock \emph{\bibinfo{journal}{SIAM Rev.}} \textbf{\bibinfo{volume}{53}},
  \bibinfo{pages}{526--543} (\bibinfo{year}{2011}).

\bibitem{Gazda2015}
\bibinfo{author}{Gazda, S.}, \bibinfo{author}{Iyer, S.},
  \bibinfo{author}{Killingback, T.}, \bibinfo{author}{Connor, R.} \&
  \bibinfo{author}{Brault, S.}
\newblock \bibinfo{title}{The importance of delineating networks by activity
  type in bottlenose dolphins (tursiops truncatus) in cedar key, florida.}
\newblock \emph{\bibinfo{journal}{R Soc Open Sci}}
  \textbf{\bibinfo{volume}{2}}, \bibinfo{pages}{140263} (\bibinfo{year}{2015}).

\bibitem{gitMVM}
\bibinfo{author}{Ortiz, E.}
\newblock \bibinfo{title}{Github: Multiscale-voter-model}.
\newblock
  \bibinfo{howpublished}{\url{https://github.com/elisendaortiz/Multiscale-Voter-Model}}
  (\bibinfo{year}{2021}).

\bibitem{Balagna2020}
\bibinfo{author}{Balagna, J.} \emph{et~al.}
\newblock \bibinfo{title}{Consensus-driven approach for decision-making in
  diverse groups}.
\newblock \emph{\bibinfo{journal}{AJPH}} \textbf{\bibinfo{volume}{110}},
  \bibinfo{pages}{5} (\bibinfo{year}{2020}).

\bibitem{McKinseyreport}
\bibinfo{author}{Hunt, V.}, \bibinfo{author}{Dixon-Fyle, S.},
  \bibinfo{author}{Prince, S.} \& \bibinfo{author}{Dolan, K.}
\newblock \emph{\bibinfo{title}{Diversity wins: How inclusion matters}}
  (\bibinfo{publisher}{McKinsey \& Company}, \bibinfo{year}{2020}).

\end{thebibliography}

\end{document}